\documentclass[11pt]{article}

\usepackage{amsmath,amssymb,amsthm,color,graphicx}

\newtheorem{theorem}{Theorem}
\newtheorem{lemma}{Lemma}
\newtheorem{claim}{Claim}
\newtheorem{proposition}{Proposition}

\newtheorem{corollary}{Corollary}
\newtheorem{observation}{Observation}

\theoremstyle{definition}
\newtheorem{definition}{Definition}

\newcommand{\bbox}{\vrule height7pt width4pt depth1pt}

\newcommand{\cell}[1]{{cell(#1)}}
\newcommand{\mangle}{\measuredangle}

\newcommand{\comment}[1]{}
\newcommand{\ignore}[1]{}

\def\clap#1{\hbox to 0pt{\hss#1\hss}}

\makeatletter
  \def\moverlay{\mathpalette\mov@rlay}
  \def\mov@rlay#1#2{\leavevmode\vtop{%
    \baselineskip\z@skip \lineskiplimit-\maxdimen
    \ialign{\hfil$#1##$\hfil\cr#2\crcr}}}
\makeatother

\def\P{{\cal P}}
\def\F{{\cal F}}

\def\B{{\cal B}}
\def\R{\mathbb{R}}

\newcommand{\remove}[1]{}

\begin{document}
\title{A generalization of Thue's theorem to packings of non-equal discs, and an  application to a discrete approximation of entropy}
\author{
Rom Pinchasi\thanks{
Mathematics Dept., Technion---Israel Institute of Technology,
Haifa 32000, Israel.
\texttt{room@math.technion.ac.il}. Supported by ISF grant (grant No. 1357/12).
}\and
Gershon Wolansky\thanks{
Mathematics Dept.,
Technion---Israel Institute of Technology,
Haifa 32000, Israel.
\texttt{gershonw@math.technion.ac.il}.
}
}

\maketitle

\begin{abstract}
In this paper we generalize the classical theorem of Thue about the optimal
circular disc packing in the plane. We are given a family of circular discs,
not necessarily of equal radii, with the property
that the inflation of every disc by a
factor of $2$ around its center does not contain any center of another disc in
the
family (notice that this implies that the family of discs is a packing).
We show that in this case the density of the given packing is at most
$\frac{\pi}{2\sqrt{3}}$, which is the density of the optimal unit disc packing.
\end{abstract}

\section{Introduction}
By a \emph{disc} we will always mean a circular disc in the plane, that is,
the set of points in the plane whose distance from some point (the center of the disc) is smaller than or equal to a number $r$, the radius of the disc.
A \emph{packing} is a family of pairwise disjoint discs in $\mathbb{R}^2$.
When a packing is contained in some bounded set $S$, then the \emph{density}
of the packing is the percentage of the volume of $S$ that is covered by the sets
of the packing. The notion of density of a packing is generalized also for
unbounded sets $S$
by exhausting them with bounded sets, usually the intersection of $S$ with larger
and larger cubes centered, say, at the origin.

The optimal unit disc packing in the plane is a celebrated classical problem that
goes back to the 1663 famous Kepler's problem about the densest unit sphere packing
in $\mathbb{R}^3$. In 1773 Lagrange proved that among all lattice packings, that is
packing of unit discs where the sets of centers is a lattice, the densest one
(the hexagonal lattice) has density $\frac{\pi}{2\sqrt{3}}$.
The densest unit disc packing problem in the plane,
without any additional assumption,
was solved only in 1910 by Thue \cite{Th10} and ever since this result is known
as Thue's theorem.
Thue's proof was considered incomplete and a full complete proof
of the theorem was given in 1943 by L. F. T\'oth \cite{T43}.
Since then more proofs, each more elegant than the other,
where presented, as this theorem and topic attracted quite some attention
(see \cite{SM44, D64, H92, CW10}).

It is not hard to see that for any bounded (open) set $S$ in the plane,
one can find a packing of discs contained in $S$, not necessarily of the same
radii, whose density is arbitrarily close to $1$. Indeed, assume we are given a
disc packing $\P$ in $S$ of density $\alpha<1$.
Consider the subset of $S$ not covered by the discs in $\P$ and decompose it into
squares covering at least half of its area. Inside each square we can place
a disc covering at least half of the area of the square.
Altogether if we add those
discs to our packing $\P$ we get another packing whose density is at least
$\alpha' > \alpha+\frac{1}{4}(1-\alpha)$.
Equivalently, $1-\alpha' < \frac{3}{4}(1-\alpha)$. We see from here that by
repeating this procedure many times we can get disc packings with density $\beta$
such that $1-\beta$ is arbitrarily close to $0$.

Therefore, the question of optimal disc packing with no further assumptions
on the family of discs in the packing is not very interesting.
There are, however, some works about packing of non-congruent discs.
In \cite{T53}, T\'oth observed that the optimal density of disc packing in the
plane remains $\frac{\pi}{2\sqrt{3}}$ even if we allow the radii of the discs
to be in the interval between $0.906$ and $1$.
This interval has been extended to $[0.702,1]$ in \cite{B69}.
Likos and Henley \cite{LH93} consider the optimal density of disc
packing that contains only discs of radii $1$ and $r$ where $r<1$ is given.
Even this, seemingly simple, problem turns to be difficult for almost all
values of $r$ with very few exceptions
(see \cite{H03}).

In this paper we consider the density of packings of circular discs with different
radii in a way that generalizes Thue's theorem on one hand and does not follow
from any of the known proofs of Thue's theorem on the other hand.

\bigskip

We say that a family $\F$ of discs in the plane is \emph{locally finite}
if every bounded set in the plane may contain only finitely many discs
in $\F$. Notice that when considering the density of a packing $\F$
there is not much loss of generality by assuming that $\F$ is locally finite.
This is because we can partition the plane into say unit squares.
In each unit square discard all the discs whose radius is small enough
so that altogether all the discarded discs do not cover more than a very small
percentage of the unit square in question. By doing this we remain with a
locally finite family of discs and the overall density
of our packing reduces only by arbitrarily small number.

We say that $\F$ has \emph{sub-linear radii growth} if
as $n$ goes to infinity the maximum radius of a disc of $\F$ contained
in a ball of radius $n$ around the origin is $o(n)$. Notice in particular that
if the radii of the discs in $\F$ are bounded then clearly $\F$ has
sub-linear radii growth. It will be convenient for us to assume that
our packing has sub-linear radii growth in order to avoid discussing
``boundary effects'' when considering the density of the packing $\F$
restricted to a large ball. We remark that for any packing $\F$, the maximum
radius of a disc in $\F$ contained in a ball $B$ of radius $n$ around the
origin (assuming $B$ contains at least two such centers, which is
true when $n$ is large enough) is clearly at most $2n$.

\begin{theorem}\label{theorem:main}
Let $\F$ be a locally finite collection of
circular discs in the plane with the property that
the inflation of every disc around its center by a factor of $2$ does not
contain any of the centers of the other discs in $\F$ (notice that such $\F$ is
necessarily a packing).
Assume that $\F$ has sub-linear radii growth. Then
the density of the packing $\F$ is not larger than the density of
the optimal unit disc packing, namely $\frac{\pi}{2\sqrt{3}}$.
\end{theorem}

Notice that the factor of $2$ in Theorem \ref{theorem:main} is best possible
and cannot be replaced by a smaller number.
Indeed, observe that a unit disc packing satisfies that condition
Theorem \ref{theorem:main}, as the distance between any two centers in a unit
disc packing is at least $2$. Therefore, Theorem \ref{theorem:main}
generalizes Thue's theorem.
If we take an optimal unit disc packing, with density $\frac{\pi}{2\sqrt{3}}$, and
inflate each disc by a factor of $1+\epsilon$ (for small positive $\epsilon$)
around its center, then the density of the union of all discs in the family
(which is not a packing anymore) is strictly greater than $\frac{\pi}{2\sqrt{3}}$.
Observe that the family of these inflated discs satisfies the condition in Theorem
\ref{theorem:main} once we replace $2$ with $\frac{2}{1+\epsilon}$.
It could be, however, that one could replace the $2$ in Theorem
\ref{theorem:main} by a smaller number, with the additional assumption that
$\F$ is a packing.

We remark that the condition in Theorem \ref{theorem:main} that
the inflation of every discs in $\F$ by a factor of $2$ does not contain a
center of any other disc in $\F$ is equivalent
to that the radius of every disc $D$ in $\F$ is at most $\frac{1}{2}$ times the
smallest distance from the center of $D$ to a center of another disc in $\F$.
We could therefore assume, without loss of generality, that for every disc
$D \in \F$ the radius of $D$ is \emph{equal} to $\frac{1}{2}$ times the
smallest distance from the center of $D$ to a center of another disc in $\F$.
\section{Entropy approximation}
In this section we introduce another motivation for
Theorem \ref{theorem:main}, which is beyond the scop of combinatorial geometry.
Here we describe the claim and sketch a proof.  The interested reader may consult \cite{CW} for a detailed discussion and further results. The non-interested  reader may skip this section, since no part of it is needed in the rest of this text.  However, to understand this section we need the definition of $Cell(D)$ corresponding to a disc $D$, as defined in the first paragraph of section \ref{sec3}.
 \par
 Consider the set ${\cal B}(\Omega)$ of  Borel probability measures on an "nice", compact set $\Omega\subset \mathbb{R}^2$ (we may assume it is a disc, or square). The {\it entropy} of a measure $\mu\in{\cal B}(\Omega)$ is defined as the Lebesgue integral
$$ E(\mu):= \int_\Omega\rho\ln\rho \ , $$
where   $\mu:=\rho dx$ if such a density exists, or $E(\mu)=+\infty$ if such a density does not exist. \par
Our object is to find a proper approximation of the entropy  on the class of $N-${\it empirical measures}:
$${\cal B}_N(\Omega):= \left\{ \mu_N=N^{-1}\sum_{i=1}^N \delta_{x_i} \ ; \ \ \ x_i\in\Omega \ , \ x_i\not= x_j \ \text{for} \ i\not= j,\ 1\leq i,j\leq N\right\}\subset {\cal B}(\Omega)$$
where $N\in \mathbb{N}$.  Let ${\cal B}_\infty(\Omega):= \cup_{N\in \mathbb{N}} {\cal B}_N(\Omega)$.
\par
Since $\Omega$ is a compact set, ${\cal B}(\Omega)$ is compact with respect to the weak  ($C_b^*(\Omega)$) topology, that is, for every sequence $\{\mu_j\}\in {\cal B}(\Omega)$ there exists a subsequence $\mu_{j_k}$ and a measure $\mu\in {\cal B}(\Omega)$ such that $\mu_{j_k}\rightharpoonup\mu$ as $k\rightarrow\infty$, that is:
$$ \lim_{k\rightarrow\infty}\int_\Omega\phi d\mu_{j_k}=\int_\Omega\phi d\mu$$
for any bounded continuous $\phi$ on $\Omega$.  It is also evident that ${\cal B}_\infty(\Omega)$ is dense in ${\cal B}(\Omega)$ with respect to the weak topology.

 Let $\Omega^\otimes:= \cup_{N\in \mathbb{N}}\Omega^{\otimes N}$ where $X\in \Omega^{\otimes N}$ iff   $X:=(x_1, \ldots x_N)$ is an {\it unordered} sequence of $N$  distinct points in $\Omega$.  Let $N(X):=N$ iff $X\in\Omega^{\otimes N}$. We first note that each $\mu\in{\cal B}_\infty(\Omega)$ can be identified with a point $X\in\Omega^{\otimes N}$.   Thus, a measure $\mu\in{\cal B}_N(\Omega)$ can be identified with $X\in\Omega^{\otimes N}$ via $\mu:= \delta_{X}\equiv N(X)^{-1}\sum_{i=1}^{N(X)} \delta_{x_i}$.
\begin{definition}\label{gamma}
A $\Gamma-$approximation of the entropy $E$ is a function ${\cal E}:\Omega^\otimes\rightarrow \R\cup\{\infty\}$ such that
\begin{description}
\item{i)} For {\it any} sequence $\{X_N\}$ such that $X_N\in\Omega^{\otimes N}$ and $\delta_{X_N}\rightharpoonup\mu\in{\cal B}(\Omega)$,
$$ \liminf_{N\rightarrow\infty} {\cal E}(X_N)\geq E(\mu) \ . $$
\item{ii)} For any $\mu\in{\cal B}(\Omega)$ such that $E(\mu)<\infty$ {\it there exists} a sequence $\{\tilde{X}_N\}\in\Omega^\otimes$ such that
$\delta_{\tilde{X}_N}\rightharpoonup\mu$ and
$$ \lim_{N\rightarrow\infty} {\cal E}(\tilde{X}_N)= E(\mu) \ . $$
\end{description}
\end{definition}
Given $X=(x_1, \ldots x_N)\in\Omega^{\otimes N}$, let $r_i(X)$ be half the minimal distance of $x_i$ to the rest of the points in $X$:
$$r_i(X):= \frac{1}{2}\min_{j\not= i}|x_i-x_j|\ . $$
We now pose the following result:
\begin{theorem}\label{th2}
$${\cal E}(X):= -\frac{2}{N(X)}\sum_{i=1}^{N(X)}\ln \left(r_i(X)\right)- \ln\left(2\sqrt{3} N(X)\right)$$
is a $\Gamma$ approximation of the entropy.
\end{theorem}
We now sketch the proof of Theorem \ref{th2}.
\par
Let $D_i(X)$ be the disc of radius $r_i(X)$ centered at $x_i$.
\par
A {\it partition rule}  $W$ is defined as a mapping between $\Omega^{\otimes}$ to a  partition of $\mathbb{R}^2$ into an essentially disjoint sets. For  $X\in\Omega^{\oplus}$  the partition rule associate a collection of $N(X)$  measurable sets $W_i:=W_i(X)\subset \R^2$ such that
\begin{description}
\item{a)} $W_i(X)\supset D_i(X)$ for any $i\in \{1, \ldots N(X)\}$
\item{b)} $\cup_{i=1}^{N(X)} W_i(X) \supseteq\Omega$.
  \item{c)} $|W_i(X)\cap W_j(X)|=0$ for any $i\not= j\in \{1, \ldots N(X)\}$. Here $|\cdot|$ stands for the Lebesgue measure of a measurable set in $\R^2$.
\item{d)} If $\{X_N\}$ is a density set in $\Omega$, i.e for any open set $U\subset\Omega$,  $\#(U\cap X_N)>0$ for all $N$ large enough, then
$\lim_{N\rightarrow\infty}\max_{i\in\{1, \ldots N\}}diameter\left( W_i(X_N)\cap\Omega\right)=0$.
\end{description}

Given a partition rule $W$ we may associate with every $\delta_X\in{\cal B}_\infty(\Omega)$  another measure $\mu\in{\cal B}(\Omega)$ which admits a density

$$  \rho^W_{X}(x):=N^{-1}(X) \sum_{i=1}^{N(X)}\frac{1_{W_j(X)\cap\Omega}(x)}{|W_j(X)\cap\Omega|} \ .  $$
Here $1_A(x)$ is the characteristic function for a set $A$, i.e $1_A(x)=1,0$ if $x\in A, x\not\in A$ respectively.

We now define the approximation entropy corresponding to the partition rule $\overrightarrow{W}$,
 $${\cal E}^W:\Omega^\otimes\rightarrow \mathbb{R}\cup\{\infty\}$$ as the entropy of $\rho^W_{X} dx$ associated with $\delta_X\in {\cal B}_\infty(\Omega)$:
\begin{equation}\label{EW} {\cal E}^W(X):= E(\rho^W_{X}dx)=\sum_{i=1}^{N(X)}\ln\left( \frac{1}{|W_i(X)\cap\Omega|}\right) -\ln N(X) \  .  \end{equation}
We show (cf. \cite{CW}):
\begin{proposition}\label{prop1}
For any partition rule $W$ verifying (a-d) above, ${\cal E}^W$ is a $\Gamma-$approximation of the  entropy.
\end{proposition}
 It is not too difficult to find partition rules.
Recall the definition of Voronoi tessellation  corresponding to $X=(x_1, \ldots)\in \Omega^{\otimes}$:
\begin{equation}\label{vor}V_i(X):= \left\{ y\in\R^2\ ; \ \ |y-x_i|\leq |y-x_j| \ \ \forall 1\leq j\leq N(X) \ \right\} \ . \end{equation}
Indeed, one can easily show that $X\rightarrow \overrightarrow{V}$ is a partition rule.
\par
We now define another partition rule: Let $cell(D_i(X))$ as defined in the first paragraph of section \ref{sec3} below. We know that $cell(D_i(X))$ are per-wise disjoint, while $\cup_{i=1}^N cell(D_i(X))\cap\Omega\subset\Omega$ with a (possibly) strict inclusion. Thus, the partition $X$ into
$\{cell(D_i(X))\}$ is {\it not} a partition rule, since it may violate condition (b) above.
\par
Let $Cell(X):=\cup_{i=1}^N cell(D_i(X))$. Define
\begin{equation}\label{mcpr} W_i(X):= cell(D_i(X))\cup\left( V_i(X)- Cell(X)\right) \ .\end{equation}
It is now easy to see that $W(X)$ so defined is a partition rule, verifying (a-d) above.
In particular, ${\cal E}^W$  as defined in (\ref{EW}) under the partition rule (\ref{mcpr}) is a $\Gamma-$approximation of the entropy, via Proposition \ref{prop1}.
\par
By the proof of Theorem \ref{theorem:main} (section \ref{sec3} below) we find out that the area of $cell(D_i(X))$ is not smaller than $ 2\sqrt{3}r^2_i(X)$, i.e  $|W_i(X)|\geq  2\sqrt{3}r^2_i(X)$.  In addition we can show  that for $X\in\Omega^\otimes$ for which $\delta_X$  approximate (in $C^*(\Omega)$) a measure $\mu\in{\cal B}(\Omega)$ satisfying $E(\mu)<\infty$,
it follows that \\
 $W_i(X)\not \subset\Omega$ for only $o(N(X))$ of the points.  So, we replace $\ln\left(|W_i(X)\cap\Omega|\right)$ in (\ref{EW}) by  $\ln\left( 2\sqrt{3}r^2_i(X)\right)$, taking advantage of the monotonicity of $\ln$, and obtain that ${\cal E}$ defined in Theorem \ref{th2} is not smaller, assymptotically, than ${\cal E}^W$ with $W$ given by (\ref{mcpr}). Thus ${\cal E}$ satisfies condition (i) of definition \ref{gamma}.
 \par
 To verify condition (ii) we recall that a optimal ratio of $|W_i(X)|/ ( 2\sqrt{3}r^2_i(X))\gtrsim 1$ is obtained for hexagonal grids.
  Given $\mu\in{\cal B}(\Omega)$, we can approximate it (in the weak topology) by a sequentially constant density. Then we construct an hexagonal grid on every domain in $\Omega$ on which this density is a constant.
   For the details of the proof see \cite{CW}.

\section{Proof of Theorem \ref{theorem:main}.}\label{sec3}
For every disc $D$ in $\F$ we define a cell, that we denote by $\cell{D}$,
in the following way. Denote by $O$ the center of $D$ and let $r$ be the
radius of $D$.
For every $D' \in \F$, different from $D$, let $O'$ be the center of $D'$
and let $r'$ denote the radius of $D'$.
Let $\ell(D,D')$ be the line perpendicular to $OO'$ such that the intersection
point $B$ of
$\ell(D,D')$ and the segment $OO'$ satisfies $\frac{|OB|}{|O'B|}=\frac{r}{r'}$.
Notice that $\ell(D,D')$ separates $D$ and $D'$ because $|OO'| \geq
\max(2r,2r') \geq r+r'$.
Let $H(D')$ denote the (open)
half-plane determined by $\ell(D,D')$ that contains $D$.
Finally, define $\displaystyle{\cell{D}=\cap_{D' \in \F \setminus \{D\}}H(D')}$.


Observe that for every disc $D \in \F$ we have $D \subset \cell{D}$ and for
any two discs $D_{1},D_{2} \in \F$ we have
$\cell{D_{1}} \cap \cell{D_{2}}=\emptyset$. If all the discs in $\F$ are of the
same radii, then the cells $\{\cell{D} \mid D \in \F\}$ are just cells of
the Voronoi diagram
of the set of centers of the discs in $\F$. For arbitrary family of discs, however,
it is possible that the collection of cells $\cell{D}$ does not cover the plane.

In order to prove Theorem \ref{theorem:main}
we restrict our attention to a large ball $\B$
around the origin. Fix $c$ to be any number strictly greater than
$\frac{\pi}{2\sqrt3}$.
We would like to show that
it is not possible to find larger and larger balls $\B$ such that the area
of $B$ contained in the union of all discs in $\F$ is more than $c$ times the
area of $\B$.

Let $\B$ be a large ball of radius $n$.
Because $\F$
has sub-linear radii growth we can discard from $\F$ all discs that are
not contained in $\B$.
This is because the union of all discs in
$\F$ intersecting the boundary of $\B$ is contained in an annulus of width
$o(n)$
whose area is $o(n^2)$ and therefore negligible
with respect to the area of $\B$.
Because $\F$ is locally finite,
$\B$ contains only finitely many discs in $\F$. We discard from $\F$ all
the discs that are not contained in $\B$.

We claim that it is enough
to show that the portion of the area of any disc $D$
in its cell $\cell{D}$ is not greater than the portion of the area of a disc
in its circumscribing hexagon, namely, $\frac{\pi}{2\sqrt3}$.
Indeed, let $r$ denote the maximum radius of a disc in $\F$
and recall that $r=o(n)$. Let $\B'$ be the ball concentric with $\B$
whose radius is equal to the radius of $\B$ plus $r$.
Add to $\F$ many more artificial discs, each with extremely small
radius, centered very densely at points on the boundary of $\B$.
Notice that $\F$ together with the additional artificial discs satisfies
the conditions in Theorem \ref{theorem:main}.
Observe that for every disc $D \in \F$ that is not artificial
the new $\cell{D}$ is a subset of the original $\cell{D}$ before the
artificial discs were added to $\F$.
Notice moreover that the new $\cell{D}$ is fully contained in $\B''$ which is
a ball concentric with $\B$ whose radius is equal to the radius of $\B$ plus
$r+1$.
Hence, if we show that every disc $D$ in $\F$ cannot cover more than
$\frac{\pi}{2\sqrt3}$ of the area of $\cell{D}$, this will show that
the union of all non-artificial discs in $\F$ cannot cover more than
$\frac{\pi}{2\sqrt3}$ of the area of $\B''$. Observe that the difference
between the area of $\B$ and the area of $\B'$ is $o(n^2)$ and
is negligible compared to the
area of $\B$ when $\B$ is a large ball (that is when $n$ is large).
This means that the union of all discs in the original family $\F$ cannot
cover from $\B$ an area of at least $c$ times the area of $\B$ for fixed
$c>\frac{\pi}{2\sqrt3}$ and a ball $\B$ that is large enough.

\bigskip

Therefore, we will concentrate on showing that
the portion of the area of any disc $D$
in its cell $\cell{D}$ (we may assume that $\cell{D}$ is bounded)
is not greater than the
portion of the area of a disc
in its circumscribing hexagon, namely, $\frac{\pi}{2\sqrt3}$.
(We note that $\frac{\pi}{2\sqrt3} \leq 0.906$.)

To this end we will show something stronger. Fix a disc $D \in \F$, denote
its center by $O$, and assume without loss of generality that it is a unit disc.
Notice that $\cell{D}$ is a convex polygon. We will show that for every edge
$e$ of $\cell{D}$ the portion of the area of $D$ inside the triangle
determined by $O$ and $e$ is at most $\frac{\pi}{2\sqrt3}$.
We further strengthen our statement as follows:
Let $C$ be the point on
the line $\ell$ through $e$ such that $OC$ is perpendicular to $\ell$.
We will show that if $A$ is a vertex of the edge $e$ such that
$AC$ overlaps with $e$, then the portion of the area of $D$ inside the
triangle $\Delta OAC$ is at most $\frac{\pi}{2\sqrt3}$
(see Figure \ref{fig:e_C}).


To see that this is indeed a stronger statement, let $A$ and $B$ be the two
vertices of the edge $e$.
We split into two possible cases.
If $C$ is a point in the segment $AB$, then both $AC$ and $BC$ overlap with
$e$. Notice that $\frac{area(D \cap \Delta OAB)}{area(\Delta OAB)}
\leq \max(\frac{area(D \cap \Delta OAC)}{area(\Delta OAC)},
\frac{area(D \cap \Delta OBC)}{area(\Delta OBC)})$.
If $C$ does not belong to the segment $AB$, then assume without loss
of generality that $B$ is a point in the segment $AC$.
We claim that
$\frac{area(D \cap \Delta OAB)}{area(\Delta OAB)} \leq
\frac{area(D \cap \Delta OAC)}{area(\Delta OAC)}$.
The reason is that the expression
$\frac{area(D \cap \Delta OAC)}{area(\Delta OAC)}$ is monotone decreasing
in the distance of $A$ from $C$, or equivalently in the angle
$\mangle AOC$ (this is because
$\frac{area(D \cap \Delta OAC)}{area(\Delta OAC)}=
\frac{\mangle AOC}{\tan \mangle AOC}$).
Therefore, $\frac{area(D \cap \Delta OAC)}{area(\Delta OAC)} \leq
\frac{area(D \cap \Delta OBC)}{area(\Delta OBC)}$.
This implies
$$
\frac{area(D \cap \Delta OAB)}{area(\Delta OAB)} =
\frac{area(D \cap \Delta OAC)-area(D \cap \Delta OBC)}
{area(\Delta OAC)-area(\Delta OAB)} \leq
\frac{area(D \cap \Delta OAC)}{area(\Delta OAC)}.
$$

We leave the verification of the last inequality to
the reader.

\bigskip

Fix an edge $e$ of $\cell{D}$ and let $D_{1}$ be the disc in $\F$ that gives
rise to the edge $e$, that is, $\ell(D,D_{1})$ contains $e$.
Denote by $A_{1}$ the point of intersection of $\ell(D,D_{1})$ and the line
through $OO_{1}$. Let $A_{2}$ be one vertex of $e$.

We have $\mangle OA_{1}A_{2} = \frac{\pi}{2}$.
Denote by $O_{1}$ the center of $D_{1}$ and denote by $r_{1}$ the radius of
$D_{1}$.
Recall, because of the definition of $\cell{D}$ and the fact that the radius of $D$ is
equal to $1$, that we have $\frac{OA_{1}}{O_{1}A_{1}}=\frac{1}{r_{1}}$.
Let $D_{2}$ be the disc in $\F$ that gives rise to the edge $e'$ of $\cell{D}$
that is adjacent to $A_{2}$ but different from $e$.
Denote by $O_{2}$ the center of $D_{2}$ and let $r_{2}$ denote the radius of
$D_{2}$ (see Figure \ref{fig:A1A2D1D2}).


For three points $A,B$, and $C$ in $\mathbb{R}^2$
we denote by $f(A,B,C)$ the ratio between the area
of $D \cap \Delta ABC$ and the area of the triangle $\Delta ABC$.
We need to show that $f(O,A_{1},A_{2}) \leq \frac{\pi}{2\sqrt{3}}$.
Assume to the contrary that $f(O,A_{1},A_{2}) > \frac{\pi}{2\sqrt{3}}$.

\begin{claim}\label{claim:0}
$\mangle A_{2}OO_{1} \leq \frac{\pi}{6}$.
\end{claim}

\noindent {\bf Proof.}
Notice that $f(O,A_{1},A_{2}) \leq
\frac{\mangle A_{2}OO_{1}}{\tan \mangle A_{2}OO_{1}}$.
It follows that we must have $\mangle A_{2}OO_{1} \leq \frac{\pi}{6}$
for otherwise $f(O,A_{1},A_{2}) \leq \frac{\pi}{2\sqrt{3}}$.
\bbox

\bigskip

The following observation follows directly from our definitions:

\begin{observation}\label{observation:ell}
Suppose $S_{1}$ and $S_{2}$ are two discs in $\F$ of radii $r_{1}$ and $r_{2}$,
respectively, and let $d$ be the distance between the centers of $S_{1}$ and
$S_{2}$. Then the distance $t$ from the center of $S_{1}$ to $\ell(S_{1},S_{2})$
is equal to $\frac{r_{1}d}{r_{1}+r_{2}}$.
\end{observation}

\noindent {\bf Proof.}
Indeed, this is because we have $\frac{d-t}{r_{2}}=\frac{t}{r_{1}}$.
\bbox

\begin{lemma}\label{lemma:alpha}
The angle $\mangle O_{2}OO_{1}$ is greater than $\frac{\pi}{6}$.
\end{lemma}

\noindent {\bf Proof.}

We will need the following observation:

\begin{claim}\label{claim:1}
Let $D'$ be a disc in $\F$ with center $O'$ and radius $r'$.
Suppose that the line $\ell(D,D')$ is at distance $1+x$
from the center $O$ of $D$. Then $r'$ satisfies
$\frac{1-x}{1+x} \leq r' \leq \frac{1+x}{1-x}$. Moreover,
the distance between the $O$ and $O'$ satisfies:
$2 \leq |OO'| \leq 2\frac{1+x}{1-x}$.
\end{claim}

\noindent {\bf Proof.}
Recall that the radius of $D$ is equal to $1$.
Denote by $t$ the distance from $O'$ to $\ell(D,D')$.
We have $\frac{t}{r'}=\frac{1+x}{1}=1+x$.
Because
$
2r' \leq OO'= t+1+x=r'(1+x)+(1+x),
$
we get
$$
r' \leq \frac{1+x}{1-x}.
$$
On the other hand we also have
$2 \leq OO'=t+1+x=r'(1+x)+(1+x)$ implying
$$
r' \geq \frac{1-x}{1+x}.
$$

\bigskip
To see the second part of the claim about the distance from $O$ to $O'$,
By our construction of $\ell(D,D')$, we have
$\frac{|OO'|-(1+x)}{r'}=\frac{1+x}{1}$. Therefore,
$|OO'|=(1+x)(1+r')$ and hence, as a consequence of the first part of the claim,
$2 \leq |OO'| \leq 2\frac{1+x}{1-x}$.
\bbox

\bigskip

We claim that
\begin{equation}\label{eq:OA_2}
|OA_{2}| \leq \frac{2}{\sqrt{3}} \leq 1.155.
\end{equation}

To see this, let $\alpha=\mangle A_{1}OA_{2}$.
By Claim \ref{claim:0}, $\alpha \leq \frac{\pi}{6}$.
We have $\frac{\pi}{2\sqrt{3}} \leq
f(O,A_{1},A_{2})=\frac{\alpha}{|OA_{2}|^2\sin \alpha \cos \alpha}$
Notice that $\frac{\alpha}{sin \alpha \cos \alpha}$ is monotone increasing
function of $\alpha$ and hence (recall $\alpha \leq \frac{\pi}{6}$)
$\frac{\pi}{2\sqrt{3}}  \leq \frac{\pi/6}{|OA_{2}|^2(1/2)(\sqrt{3}/2)}$,
implying (\ref{eq:OA_2}).

As a consequence of (\ref{eq:OA_2}), the distance from $O$ to $\ell(D,D_{2})$
is at most $1.155$. Moreover, the distance from $O$ to $A_{1}$ is smaller than
the distance from $O$ to $A_{2}$ and therefore we also deduce that the
distance from $O$ to $\ell(D,D_{1})$ is at most $1.155$.
By Claim \ref{claim:1}, both distances from $O$ to $O_{1}$ and from
$O$ to $O_{2}$ are at least $2$ and at most $2\frac{1+0.155}{1-0.155}<2.74$.

By Claim \ref{claim:1} and the fact that the distance from $O$ to both
$\ell(D,D_{1})$ and $\ell(D,D_{2})$ is at most $1.155$, we have that
both $r_{1}$ and $r_{2}$ are at least $\frac{1-0.155}{1+0.155} \geq 0.73$.
Hence $|O_{1}O_{2}| \geq r_{1}+r_{2} \geq 1.46$.
Because $2 \leq |OO_{1}|,|OO_{2}| \leq 2.74$ there are two extreme options.
In one $\mangle O_{1}OO_{2}$ is at least as large as the angle at
$P$ in a triangle $\Delta PQR$ such that $|PQ|=2.74$, $|QR|=1.46$, and
$|PR|=2$.
The cosine of this angle $a$ satisfies
$\cos a=\frac{2.74^2+2^2-1.46^2}{2\cdot2 \cdot 2.74} \leq 0.856 <
\cos \frac{\pi}{6}$.

The other extreme case is where
$\mangle O_{1}OO_{2}$ is at least as large as the angle at
$P$ in a triangle $\Delta PQR$ such that $|PQ|=|PR|=2.74$ and $|QR|=1.46$.
The cosine of this angle $a$ satisfies
$\cos a=\frac{2.74^2+2.74^2-1.46^2}{2\cdot 2.74 \cdot 2.74} \leq 0.859 <
\cos \frac{\pi}{6}$.
\bbox

\bigskip

Because $\mangle O_{2}OO_{1}$ is the angle generated between
$\ell(D,D_{2})$ and (the right ray of) $\ell(D,D_{1})$ we have the following

\begin{corollary}\label{cor:alpha}
The angle generated between $\ell(D,D_{2})$ and (the right ray of)
$\ell(D,D_{1})$ is greater than $\frac{\pi}{6}$.
\end{corollary}

The next lemma will turn to be quite useful.

\begin{lemma}\label{lemma:lDD1}
Let $\ell$ be a line parallel to $\ell(D,D_{1})$ that separates
$\ell(D,D_{1})$ and $D$. In particular, $\ell$ is closer to $O$ than
$\ell(D,D_{1})$ is. Let $A^{*}_{1}$ be the intersection point of
$\ell$ with the line through $O$ and $O_{1}$ and let $A^*_{2}$ be the
intersection point of $\ell$ with the line $\ell(D,D_{2})$.
Then $f(O,A_{1},A_{2}) \leq f(O,A^*_{1},A^*_{2})$.
\end{lemma}

\noindent {\bf Proof.}
Denote by $B$ the intersection point of $\ell(D_{2},D)$ and the line through
$O$ and $O_{1}$.
Let $\alpha$ denote the measure of the angle between $\ell(D_{2},D)$ and the
(positive part of the) $x$-axis, that is, $\alpha=\mangle BA_{2}A_{1}$.
Denote by $x$ the angle $\mangle A_{2}OA_{1}$ (see Figure \ref{fig:criticalD1}).


It is not hard to express $f(O,A_{1},A_{2})$ as a function of $x$:
We have
$f(O,A_{1},A_{2}) = \frac{x/2}{\mbox{area of~$\Delta OA_{1}A_{2}$}}$.

Notice that from the theorem of sines
$|A_{2}O|=\frac{|OB|\sin (\pi/2-\alpha)}{\sin (\pi/2-x+\alpha)}=
\frac{|OB|\cos \alpha}{\cos (x-\alpha)}$.
Therefore, the area of $\Delta OA_{1}A_{2}$ is equal to
$$
\frac{1}{2}|OA_{1}||A_{1}A_{2}|=\frac{1}{2}
\sin x \cos x |OA_{2}|^2=\frac{1}{2}
\sin x \cos x \frac{|OB|^2\cos^2 \alpha}{\cos^2 (x-\alpha)}=
\frac{1}{2}\sin x \cos x \frac{|OB|^2\cos^2 \alpha}{(1+\cos (2x-2\alpha))/2}.
$$

Since $f(O,A_{1},A_{2}) = \frac{x/2}{\mbox{area of~$\Delta OA_{1}A_{2}$}}$,
then
up to positive constant multipliers that depend only on $B$, $O$, and $\alpha$,
this function is equal to $g(x)=\frac{x(1+\cos(2x-2\alpha))}{\sin 2x}$.

Let $x'$ denote the angle $\mangle A^*_{2}OA^*_{1}$
Notice that $x'>x$.
Therefore, in order to show that $f(O,A_{1},A_{2}) \leq f(O,A^*_{1},A^*_{2})$
it is enough to show that the function $g(x)$
is an increasing function of $x$, or equivalently that $g'(x) \geq 0$.

A direct attempt to prove $g'(x) \geq 0$ leads to the equivalent inequality
$$
\sin 2x (1+\cos (2x-2\alpha)-x2\sin(2x-2\alpha))  \geq
x(1+\cos(2x-2\alpha))2\cos2x.
$$

This reduces to
$$
(1+\cos(2x-2\alpha))\sin2x \geq x(2\cos 2x +2\cos 2\alpha)
$$
and then to
$$
\cos (x-\alpha)\sin 2x \geq 2x\cos(x+\alpha).
$$

Using the fact that $\frac{\sin 2x}{2x} \geq \cos 2x$, it will be enough
to show that

$$
\cos (x-\alpha)\cos 2x \geq \cos (x+\alpha).
$$

This is equivalent to
$$
\frac{1}{2}(\cos (3x-\alpha) + \cos (x+\alpha)) \geq \cos (x+\alpha).
$$

This finally reduces to
$$
\cos (3x-\alpha) \geq \cos (x+\alpha)
$$
which is equivalent to $x \leq \alpha$. This last inequality holds because
we have $x \leq \frac{\pi}{6} \leq \alpha$ (the first inequality is Claim
\ref{claim:0} and the second inequality is by
Corollary \ref{cor:alpha}).
\bbox

\bigskip

\begin{lemma}\label{lemma:p3}
$\mangle O_{2}OO_{1} \leq \frac{\pi}{3}$.
\end{lemma}

\noindent {\bf Proof.}
We will show that if $\mangle O_{2}OO_{1} > \frac{\pi}{3}$, then
$f(O,A_{1},A_{2}) \leq \frac{\pi}{2\sqrt{3}}$.
By Lemma \ref{lemma:lDD1}, it is enough to consider the case where
$\ell(D,D_{1})$ is tangent to $D$. In this case notice that
if $\mangle O_{2}OO_{1} = \frac{\pi}{3}$, then
$f(O,A_{1},A_{2}) = \frac{\pi}{2\sqrt{3}}$. As we further increase
$\mangle O_{2}OO_{1}$, the value of $f(O,A_{1},A_{2})$ decreases.
\bbox


\bigskip

\noindent {\Large \bf Reducing to the critical case}

We say that $D$ is \emph{critical} if
its radius, namely $1$, is equal to $\frac{1}{2}\min(|OO_{1}|,|OO_{2}|)$.
Intuitively speaking, we inflate $D$ around its center
as much as we can so that the conditions in Theorem \ref{theorem:main}
are still satisfied when restricting our attention only to the three discs
$D$, $D_{1}$, and $D_{2}$.

In a similar way we define the notion of critical for $D_{1}$ and $D_{2}$.
That is, $D_{1}$ is critical if $r_{1}$ is equal to
$\frac{1}{2}\min(|O_{1}O|,|O_{1}O_{2}|)$.
We say that $D_{2}$ is critical if
$r_{2}=\frac{1}{2}\min(|O_{2}O|,|O_{2}O_{1}|)$.

In this subsection we will show that one can assume,
without loss of generality,
that all three discs $D,D_{1}$, and $D_{2}$ are critical. This reduction
will simplify a lot the presentation of the rest of the proof.

Without loss of generality we will assume that $O$ is the origin, $O_{1}$
lies strictly above $O$ on the $y$-axis and $O_{2}$ lies in the half-plane
$\{x<0\}$.

It is easiest to see that we may assume that $D_{2}$ is critical.
Indeed, by increasing the value of $r_{2}$ we push the line $\ell(D_{2},D)$
towards $O$, thus shifting the point $A_{2}$ to the right. This increases
the value of $f(O, A_{1},A_{2})$ (see Figure \ref{fig:criticalD2}).
Formally, denote by $x$ the angle
$\mangle A_{2}OA_{1}$. We have
$f(O,A_{1},A_{2})=\frac{x/2}{\frac{1}{2}\tan x |OA_{1}|}$. This is a decreasing
function of $x$. Hence, as $A_{2}$ moves to the right $x$ decreases and
consequently $f(O,A_{1},A_{2})$ increases.


\bigskip

Next, we claim that we may assume without loss of generality that $D_{1}$ is
critical.
To see this notice that
as we increase $r_{1}$, we push the line $\ell(D,D_{1})$ towards $O$
(this operation has an effect both on $A_{1}$ and on $A_{2}$).
By Lemma \ref{lemma:lDD1}, as we push the line $\ell(D,D_{1})$ towards $O$,
the value of $f(O,A_{1},A_{2})$ does not decrease.

\bigskip

Finally, we claim that
we may assume without loss of generality that $D$ is critical.
To see this.
we will now show that the effect of increasing the radius of $D$ is equivalent
to keeping $D$ a unit disc and pushing the lines $\ell(D,D_{1})$ and
$\ell(D,D_{2})$ closer to $O$. Once we show this then the claim follows
from Lemma \ref{lemma:lDD1} because it is shown there
that pushing $\ell(D,D_{1})$ closer to $O$ (keeping $\ell(D,D_{2})$ fixed)
increases $f(O,A_{1},A_{2})$. If in addition we also push $\ell(D,D_{2})$
closer to $O$, then $f(O,A_{1},A_{2})$ can only further increase.

To see the effect of increasing the radius of $D$, let $D'$ be any other disc
in $\F$ and let $O'$ and $r'$ be its center and radius, respectively.
The distance $d$ from $O$ to $\ell(D,D')$ satisfies
$\frac{1}{d}=\frac{r'}{|OO'|-d}$, namely, $d=\frac{|OO'|}{1+r'}$.
If we increase the radius of $D$ to be $r>1$, then
the new distance $d'$ from $O$ to $\ell(D,D')$ satisfies
$\frac{r}{d'}=\frac{r'}{|OO'|-d'}$, namely, $d'=\frac{|OO'|}{1+r'/r}$.
Scaling back the picture so that $D$ is again a unit disc, this distance
reduces to $\frac{1}{r}\frac{|OO'|}{1+r'/r}=\frac{|OO'|}{r+r'}$.
Because $r>1$ we have $\frac{|OO'|}{r+r'} < \frac{|OO'|}{1+r'}=d$.


\noindent {\Large \bf Concluding the proof}

We henceforth assume that all three discs $D,D_{1},$ and $D_{2}$ are critical.
We split into three cases according to which is the closest pair of centers
among $O,O_{1},$ and $O_{2}$.

\noindent {\bf Case 1.} $|OO_{1}| \leq |OO_{2}|, |O_{1}O_{2}|$.
In this case, because $D$ and $D_{1}$ are both critical,
the radii of both $D$ and $D_{1}$ are the same
and are equal to $\frac{|OO_{1}|}{2}$.
As we assume that $D$ is a unit disc,
the radii of both $D$ and $D_{1}$ are equal to $1$ and hence $|OO_{1}|=2$.
The discs $D$ and $D_{1}$ touch each other at $A_{1}$
and $\ell(D,D_{1})$ is their common tangent at $A_{1}$.
Let $A_{2}'$ be the point on $\ell(D,D_{1})$ to the left of $A_{1}$ such that
$f(O,A_{1},A_{2}')=\frac{\pi}{2\sqrt3}$.
In order for $f(O,A_{1},A_{2})$ to be greater than
$\frac{\pi}{2\sqrt3}$ the line $\ell(D,D_{2})$
must cross $\ell(D,D_{1})$ at a point $A_{2}$ to the right of $A_{2}'$.
Let $D_{1}'$ be the disc centered at $O_{1}$ whose radius is $2$ (double
the radius of $D_{1}$). The point $O_{2}$, the center of $D_{2}$,
must lie outside $D_{1}'$
because of the assumptions in Theorem \ref{theorem:main}.


We will now show that we may assume without loss of generality that
$O_{2}$ lies on the boundary of $D_{1}'$.
Let $O_{2}'$ denote the intersection point of the line through $O$ and $O_{2}$
with the boundary of $D_{1}'$.
We will replace $D_{2}$ with $D'_{2}$, the disc of radius $1$ centered at
$O'_{2}$.
By Lemma \ref{lemma:p3}, $\mangle O_{2}OO_{1} < \frac{\pi}{3}$.
This implies that both points
$O_{2}$ and $O'_{2}$
are closer to $O_{1}$ than to $O$.
We will show that $O$ is closer to the line
$\ell(D,D_{2}')$ than to the line $\ell(D,D_{2})$ (see Figure \ref{fig:case1}).
This will imply that by replacing $D_{2}$ with $D'_{2}$ we push $A_{2}$
further to the right (on $\ell(D,D_{1})$)
and therefore can only increase $f(O,A_{1},A_{2})$.

Recall that as $D_{2}$ is critical then $r_{2}$, the radius of $D_{2}$,
is equal to $\frac{1}{2}|O_{2}O_{1}|$ (this is because
$|O_{2}O_{1}| \leq |O_{2}O|$ and $r_{2}=\frac{1}{2}\min(|O_{2}O|, |O_{2}O_{1}|)$).

The distance $d$ from $O$ to $\ell(D,D_{2})$ satisfies
$\frac{1}{d}=\frac{r_{2}}{|OO_{2}|-d}$. Therefore,
$d=\frac{|OO_{2}|}{r_{2}+1}=\frac{|OO_{2}|}{\frac{1}{2}|O_{2}O_{1}|+1}$.

The distance $d'$ from $O$ to $\ell(D,D_{2}')$ satisfies
$d'=\frac{1}{2}|OO_{2}'|$ (this is because both $D$ and $D_{2}'$ are unit discs
and therefore $\ell(D,D_{2}')$ is the perpendicular bisector of $OO_{2}'$).

We claim that $d' \leq d$, or equivalently,
$$
\frac{1}{2}|OO_{2}'| \leq \frac{|OO_{2}|}{\frac{1}{2}|O_{2}O_{1}|+1}.
$$

After dividing by $2$ we get
$$
\frac{|OO_{2}'|}{4} \leq \frac{|OO_{2}|}{|O_{2}O_{1}|+2}.
$$

Keeping in mind that $2=|OO_{1}|$ and $4=|OO_{1}|+|O'_{2}O_{1}|$, we need to
show that
\begin{equation}\label{eq:case1}
\frac{|OO_{2}'|}{|OO_{1}|+|O_{2}'O_{1}|} \leq
\frac{|OO_{2}|}{|O_{2}O_{1}|+|OO_{1}|}.
\end{equation}

Notice that
$$
\frac{|OO_{2}|}{|O_{2}O_{1}|+|OO_{1}|}=
\frac{|OO'_{2}|+|O'_{2}O_{2}|}{|O_{2}O_{1}|+|OO_{1}|} \geq
\frac{|OO'_{2}|+|O'_{2}O_{2}|}{|O'_{2}O_{1}|+|O'_{2}O_{2}|+|OO_{1}|}.
$$

Hence, in order to show (\ref{eq:case1}) it is enough to show
$$
\frac{|OO_{2}'|}{|OO_{1}|+|O_{2}'O_{1}|} \leq
\frac{|OO'_{2}|+|O'_{2}O_{2}|}{|O'_{2}O_{1}|+|O'_{2}O_{2}|+|OO_{1}|}.
$$
This last inequality reduces, after elementary manipulations, to the triangle
inequality $|OO'_{2}| \leq |OO_{1}|+|O'_{2}O_{1}|$.

\bigskip

Therefore, we assume that the center $O_{2}$ of $D_{2}$ is on the boundary of
$D_{1}'$ and that the radius of $D_{2}$ is equal to $1$ (as $D_{2}$
can be assumed to be critical). Now it is easy to see that
$\ell(D_{2},D)$ passes through
$O_{1}$ and therefore it intersects with $\ell(D,D_{1})$ (at the point $A_{2}$)
to the left of $A_{2}'$
and not as required. Hence $f(O,A_{1},A_{2}) \leq \frac{\pi}{2\sqrt{3}}$.

\bigskip

\noindent {\bf Case 2.} $|OO_{2}| \leq |OO_{1}|, |O_{1},O_{2}|$.
In this case the radii of both $D$ and $D_{2}$ are equal,
and therefore are equal to $1$,
which, in turn, is half of the distance from $O$ to $O_{2}$.
Moreover, the discs $D$ and $D_{2}$ touch each other
and $\ell(D,D_{2})$ is their common tangent at the point where they touch.


By Lemma \ref{lemma:lDD1}, $\mangle O_{2}OO_{1} \leq \frac{\pi}{3}$.
Similar to the argument in Case 1, we let $D_{2}'$ denote the disc of radius $2$
centered at $O_{2}$. Observe that $O_{1}$ must be outside $D'_{2}$. This,
together with the fact that $\mangle O_{2}OO_{1} \leq \frac{\pi}{3}$, implies
that
$|O_{1}O_{2}| \leq |O_{1}O|$. This is equivalent to saying that $O_{1}$ and $O_{2}$
lie in the same half-plane bounded by $\ell(D,D_{2})$.



Let $O_{1}'$ be the intersection point of the line
through $O$ and $O_{1}$ with the boundary of $D_{2}'$.
Let $D'_{1}$ be the unit disc centered at $O'_{1}$.

\begin{claim}\label{claim:case2}
The distance from $O$ to $\ell(D,D_{1})$ is greater than or equal to
the distance from $O$ to $\ell(D,D'_{1})$.
\end{claim}

\noindent {\bf Proof.}
Let $x$ denote $\mangle OO_{1}O_{2}$ and let $\alpha = \mangle O_{1}OO_{2}$.
We will now express the distance
from $O$ to $\ell(D,D_{1})$ as a function of $x$.
By Observation \ref{observation:ell}, the distance from $O$ to $\ell(D,D_{1})$
is equal to $\frac{|OO_{1}|}{r_{1}+1}=\frac{|OO_{1}|}{|O_{1}O_{2}|/2+1}$
(here $r_{1}=|O_{1}O_{2}|/2$ because $D_{1}$ is critical and
$|O_{1}O_{2}| \leq |O_{1}O|$).
From the theorem of sines with respect to triangle $\Delta OO_{1}O_{2}$,
$|O_{1}O_{2}|=\frac{2\sin \alpha}{\sin x}$ and
$|OO_{1}|=\frac{2\sin (\alpha+x)}{\sin x}$.


Therefore, the distance from $O$ to $\ell(D,D_{1})$
is equal to $\frac{2\sin(\alpha+x)}{\sin \alpha+\sin x}$.
By checking the derivative of this function with respect to $x$ one can see
that this function is monotone decreasing in $x$.
Because $\mangle OO_{1}O_{2} \leq \mangle OO_{1}'O_{2}$, this shows that
the distance from $O$ to $\ell(D,D'_{1})$ is smaller than the distance
from $O$ to $\ell(D,D_{1})$.
\bbox

Hence by taking $D_{1}=D'_{1}$ we push $\ell(D,D_{1})$ closer to $O$
and therefore, by Lemma \ref{lemma:lDD1}, we increase the value
of $f(O,A_{1},A_{2})$.

Finally, observe that when $D_{1}=D'_{1}$ the line $\ell(D,D_{1})$ passes
through
$O_{2}$.
Let $x= \mangle A_{1}OA_{2}$ and notice that $x$ is monotone decreasing
in $\mangle O_{1}O_{2}O$ while $|OA_{2}|$ is monotone increasing in
$\mangle O_{1}O_{2}O$ (see Figure \ref{fig:case22}).


We have
$$
f(O,A_{1},A_{2})=\frac{x/2}{\frac{1}{2}|OA_{2}|^2\sin x \cos x}=
\frac{x}{\frac{1}{2}|OA_{2}|^2\sin 2x}.
$$
Therefore, as $\mangle O_{1}O_{2}O$ increases $x$ decreases
and so $\frac{x}{\sin2x}$ decreases (as can be easily verified this is
an increasing function of $x$). On the top of this $|OA_{2}|$ increases and
hence $f(O,A_{1},A_{2})=\frac{x}{\sin2x}\frac{2}{|OA_{2}|^2}$ decreases.

Notice that $\mangle O_{1}O_{2}O$ is minimum when $D, D_{2}$ and $D_{1}$
(which is now equal to $D'_{1}$) are three pairwise touching unit discs.
In the latter case we have $f(O,A_{1},A_{2})=\frac{\pi}{2\sqrt{3}}$,
showing that indeed
$f(O,A_{1},A_{2}) \leq \frac{\pi}{2\sqrt{3}}$ in general.

\bigskip

\noindent {\bf Case 3.} $|O_{1}O_{2}| \leq |OO_{1}|, |OO_{2}|$.
In this case the radii of both $D_{1}$ and $D_{2}$ are equal
and $D_{1}$ and $D_{2}$ touch each other. Denote by $r$ the radii of
$D_{1}$ and $D_{2}$. Notice that $r \leq 1$. This is because $D$ is critical
and therefore either $|OO_{1}|=2$, or $|OO_{2}|=2$ and in either case
we have $2r \leq 2$.
Because of the assumption in Theorem \ref{theorem:main}
both $|OO_{1}|$ and $|OO_{2}|$ are greater than or equal to twice the radius
of $D$, namely $2$.
We split into two sub-cases according to which of $|OO_{1}|$ and $|OO_{2}|$
is equal to $2$.

\noindent {\bf Subcase a.} $|OO_{1}|=2$ and $|OO_{2}| \geq 2$.
We claim that we may assume that $|OO_{2}|=2$. To see this we rotate
the disc $D_{2}$ around the center $O_{1}$ of $D_{1}$ in the clockwise direction
until $|OO_{2}|=2$ and we keep track of $f(O,A_{1},A_{2})$.
Notice that by rotating the disc $D_{2}$ around $O_{1}$ we only change the
position of $A_{2}$ while $O$ and $A_{1}$ remain fixed.

Let $\alpha$ denote the angle $\mangle O_{2}OO_{1}$ and notice that
as we rotate $D_{2}$ clockwise around $O_{1}$ until $|O_{2}O|=2$
$\alpha$ increases.

We have $|O_{1}O_{2}|=2r$, $|OO_{1}|=2$ and it is not hard to see that
$|OO_{2}|=2(\cos \alpha+\sqrt{r^2-\sin^2 \alpha})$. Notice that $|OO_{2}|$
is a monotone decreasing function of $\alpha$.

By Observation \ref{observation:ell}, the distance from $O$ to $\ell(D,D_{1})$
is equal to $\frac{2}{r+1}$. The distance from $O$ to $\ell(D,D_{2})$
is equal to $\frac{|OO_{2}|}{r+1}$.
Let $x$ denote the angle $\mangle A_{2}OA_{1}$. We have
$\cos x = \frac{|OA_{1}|}{|OA_{2}|}$.
Let $B$ denote the intersection point of $\ell(D,D_{2})$ and the line $OO_{2}$
(see Figure \ref{fig:case3a}).


Recall that $|OB|$, the distance from $O$ to $\ell(D,D_{2})$ is equal to
$\frac{|OO_{2}|}{r+1}$. We have $\cos (\alpha-x)=\frac{|OB|}{|OA_{2}|}$.

Hence
$$
\frac{\cos x}{|OA_{1}|}=\frac{\cos (\alpha-x)}{|OB|}=
\frac{\cos\alpha \cos x + \sin \alpha \sin x}{|OB|}.
$$
From here we conclude that
\begin{equation}\label{eq:case3a}
|A_{2}A_{1}|=|OA_{1}|\tan x=
\frac{|OB|}{\sin \alpha}-\frac{|OA_{1}|\cos \alpha}{\sin \alpha}.
\end{equation}

It will therefore be enough to show that the right hand side of
(\ref{eq:case3a}) decreases as
we increase $\alpha$. Keeping in mind that $|OA_{1}|=\frac{2}{r+1}$
and
$$
|OB|=\frac{|OO_{2}|}{r+1}=
\frac{2(\cos \alpha+\sqrt{r^2-\sin^2 \alpha})}
{r+1},
$$

The right hand side of (\ref{eq:case3a}) becomes
$$
\frac{2}{r+1}\sqrt{\frac{r^2}{\sin^2 \alpha}-1},
$$
which is evidently a decreasing function of $\alpha$.

\bigskip

We conclude that we may assume in Subcase a of Case 3
that $|OO_{1}|=|OO_{2}|=2$. Let $\alpha$ denote the angle
$\mangle A_{1}OA_{2}$. Notice that $\alpha$ is a monotone increasing function
of $r$ the radii of both $D_{1}$ and $D_{2}$.
We will show that $f(O,A_{1},A_{2})$ is an increasing function of $\alpha$.
From this it will follow that one can assume that $r$ is maximum possible,
namely $r=1$, but in this case $f(O,A_{1},A_{2})=\frac{\pi}{2\sqrt{3}}$, as
can be easily seen.

Notice that $r=2\sin \alpha$ and therefore
$|OA_{1}|=\frac{2}{r+1}=\frac{2}{1+2\sin \alpha}$.

We have
$$
f(O,A_{1},A_{2})=\frac{\alpha}{|OA_{1}||OA_{2}|}=
\frac{\alpha}{\tan \alpha |OA_{1}|^2}=
\frac{\alpha (1+2\sin \alpha)^2}{4 \tan \alpha}.
$$

It remains to show that this is an increasing function of $\alpha$.
Considering the derivative of this function, it is equivalent to showing that
$$
\frac{\alpha}{\sin \alpha} \leq
\cos \alpha \frac{1+2\sin \alpha}{1+2\sin \alpha-4\sin \alpha \cos^2\alpha}.
$$

As $\frac{\alpha}{\sin \alpha} \leq \frac{1}{\cos \alpha}$ for every
$0 \leq \alpha < \frac{\pi}{2}$, it is enough to show that

\begin{equation}\label{eq:cos}
\frac{1}{\cos \alpha} \leq
\cos \alpha \frac{1+2\sin \alpha}{1+2\sin \alpha-4\sin \alpha \cos^2\alpha}.
\end{equation}

The reduces, after elementary manipulations, to

\begin{equation}\label{eq:cos2}
\cos^2 \alpha +\sin 2\alpha \cos \alpha +2\sin\alpha \cos 2\alpha \geq
1
\end{equation}

This clearly holds for every $\alpha \leq \frac{\pi}{6}$ (which we assume)
because for those $\alpha$ we have $\cos 2\alpha \geq \sin \alpha$
and therefore the left hand side of (\ref{eq:cos2}) is at least
$\cos^2 \alpha + \sin^2 \alpha$, that is, at least $1$.

\noindent {\bf Subcase b.} $|OO_{2}|=2$ and $|OO_{1}| \geq 2$.
We claim that we may assume in this case that $r_{1}=r_{2}=1$. This will imply
$|O_{1}O_{2}|=|OO_{2}| \leq |OO_{1}|$ and we may
thus reduce to Case 2.
To see that we may assume $r_{1}=r_{2}=1$,
we will increase the value of $r$ keeping the angle $\mangle OO_{2}O_{1}$,
that we denote by $\beta$, fixed. Through this increment
we will keep $D_{1}$ and $D_{2}$ touching each other.
At every moment denote by $\alpha$ the angle $\mangle O_{2}OO_{1}$
and notice that $\alpha$ is a monotone increasing function of $r$.
We will show that as $\alpha$ increases the value of $f(O,A_{1},A_{2})$
increases.

Let $B$ denote the intersection point of $\ell(D,D_{2})$ with the line
$OO_{2}$. As $|OO_{2}|=2$, it follows from Observation \ref{observation:ell}
that $|OB|=\frac{2}{r+1}$.
By considering the triangle $\Delta OO_{2}O_{1}$ and using the theorem
of sines, we see that $|OO_{1}|=\frac{2\sin \beta}{\sin (\alpha+\beta)}$.
Therefore, again by Observation \ref{observation:ell}, we have
$|OA_{1}|=\frac{|OO_{1}|}{r+1}=\frac{2\sin \beta}{(r+1)\sin (\alpha+\beta)}$.
Recall that $A_{2}$ is the intersection point of $\ell(D,D_{2})$ and
$\ell(D,D_{1})$. Denote by $x$ the angle $\mangle A_{2}OA_{1}$ and notice that
$f(O,A_{1},A_{2})=\frac{x}{\tan x |OA_{1}|^2}$ (see Figure \ref{fig:case3b}).


By considering triangle $\Delta OA_{2}A_{1}$, we see that
\begin{equation}\label{eq:111}
\frac{1}{|OA_{2}|}=\frac{\cos x}{|OA_{1}|}=
\frac{\cos x (r+1)\sin (\alpha+\beta)}{2\sin \beta}.
\end{equation}
By considering the triangle $\Delta OA_{2}B$, we see that
\begin{equation}\label{eq:222}
\frac{1}{|OA_{2}|}=\frac{\cos (\alpha-x)}{|OB|}=
\frac{(r+1)\cos (\alpha-x)}{2}.
\end{equation}

From (\ref{eq:111}) and (\ref{eq:222}) it follows that

$$
\frac{\cos x \sin (\alpha+\beta)}{\sin \beta}=
\cos (\alpha-x)=\cos \alpha \cos x + \sin \alpha \sin x.
$$

This implies

$$
\frac{\sin x}{\cos x}=\frac{\cos \beta}{\sin \beta}.
$$

This means that the angle $x$ remains fixed through the increment of the value
of $r$ and therefore, in order to show that the
value of $f(O,A_{1},A_{2})$ increases it is enough to show that
$|OA_{1}|$ decreases, because $f(O,A_{1},A_{2})=\frac{x}{\tan x |OA_{1}|^2}$.
To see that the value of $|OA_{1}|$ decreases as we increase $\alpha$, we
recall that
$|OA_{1}|=\frac{2\sin \beta}{(r+1)\sin (\alpha+\beta)}$ and therefore it
is enough to show that $(r+1)\sin (\alpha+\beta)$ increases as we increase
$\alpha$. To this end consider triangle $\Delta O_{1}OO_{2}$
and use the theorem of sines to see that
$\frac{2r}{\sin \alpha}=\frac{2}{\sin (\alpha+\beta)}$. This implies
$r=\frac{\sin \alpha}{\sin (\alpha+\beta)}$. Using this, we see that
$$
(r+1)\sin (\alpha+\beta)=\sin \alpha + \sin (\alpha+\beta)=
2\sin(\alpha + \frac{1}{2}\beta)\cos(\frac{\beta}{2}).
$$
Now, it is enough to observe that
$0 \leq \alpha + \frac{1}{2}\beta \leq \pi/2$.
This is because $\alpha+ \beta + \alpha \leq
\alpha+ \beta + \mangle O_{2}O_{1}O = \pi$.

\bigskip
\bigskip

\bbox

\end{document}